\documentclass[12pt]{article}
\usepackage{amssymb,amsmath}
\usepackage[noblocks]{authblk}
\usepackage[top=0.75in, bottom=0.75in, left=0.75in, right=0.75in, dvips]{geometry}
\usepackage{caption}
\usepackage{graphicx}
\usepackage{epstopdf}
\begin{document}
\textwidth 10.0in
\textheight 9.0in
\topmargin -0.60in
\title{Radiative Corrections in 2+1 Dimensional Supergravity}
\author[1,2]{D.G.C. McKeon}
\affil[1] {Department of Applied Mathematics, The
University of Western Ontario, London, ON N6A 5B7, Canada}
\affil[2] {Department of Mathematics and
Computer Science, Algoma University,\newline Sault Ste. Marie, ON P6A
2G4, Canada}  

\maketitle                                 
   
\maketitle
\noindent
email: dgmckeo2@uwo.ca

\begin{abstract}
Supergravity in $2+1$ dimensions has a set of first class constraints that result in two Bosonic and one Fermionic gauge invariances.  When one uses Faddeev-Popov quantization, these gauge invariances result in four Fermionic scalar ghosts and two Bosonic Majorana spinor ghosts. The BRST invariance of the effective Lagrangian is found.  As an example of a radiative correction, we compute the phase of the one-loop effective action in the presence of a background spin connection, and show that it vanishes.  This indicates that unlike a spinor coupled to a gauge field in $2+1$ dimensions, there is no dynamical generation of a topological mass in this model.  An additional example of how a BRST invariant effective action can arise in a gauge theory is provided in an appendix where the BRST effective action for the classical Palatini action in $1 + 1$ dimensions is examined.
\end{abstract}

The first-order action for supergravity in $2+1$ dimensions is [1]
\begin{equation}
\mathcal{L}_{c1} = \epsilon^{\mu\nu\lambda} \left( b_\mu^i R_{\nu\lambda i} (w) + \overline{\psi}_\mu D_\nu \psi_\lambda\right).
\end{equation}
It can be shown that the first class constraints present in this model lead to the following gauge invariances [2]\
\begin{subequations}
\begin{align}
\delta b_\mu^i &= - \left[ \mathcal{D}_\mu^{ij} A_j - \frac{1}{2} \epsilon^{ijk}b_{\mu j} B_k + \frac{i}{4} \overline{C}\gamma^i \psi_\mu\right]\\
\delta w_\mu^i &= - \frac{1}{2} \mathcal{D}_\mu^{ij} B_j\\
\delta \psi_\mu &= - \frac{1}{2} \left(D_\mu C -\frac{i}{2} B \cdot \gamma \psi_\mu\right).
\end{align}
\end{subequations}
(In eqs. (1,2), $b_\mu^i$, $w_\mu^i$, $A_i$, $B_i$ are not Grassman (BE) and $\psi_\mu$, $C$ are Majorana-Grassmann (FD) spinors.)

If we use the gauge fixing conditions
\begin{equation}\tag{3a-c}
\partial^\mu b_\mu^i = \partial^\mu w_\mu^i = \partial^\mu\psi_\mu = 0
\end{equation}
in order to eliminate the redundant fields occurring because of the gauge invariances of eq. (2), then the gauge fixing Lagrangian is
\begin{equation}\tag{4}
\mathcal{L}_{gf} = \frac{\alpha_b}{2}N_b^i N_{bi} + \frac{\alpha_\omega}{2} N_w^iN_{wi} + \frac{\alpha_\psi}{2}\overline{N}_\psi N_\psi - N_b^i \partial^\mu b_{\mu i} - N_w^i \partial^\mu w_{\mu i} - \overline{N}_\psi \partial^\mu \psi_\mu 
\end{equation}
where $N_b^i$ and $N_w^i$ are BE while $N_\psi$ is a FD Majorana spinor.  The Faddeev-Popov (FP) quantization procedure leads to the ghost action
\begin{equation}
\hspace{-3cm}\mathcal{L}_{FP} = e_i \partial^\mu  \left(  - \mathcal{D}_\mu^{ij} c_j + \frac{1}{2} \epsilon^{ipj} b_{\mu p} + \frac{1}{4} \overline{\psi}_\mu \gamma^i \zeta \right) \nonumber 
\end{equation}
\begin{equation}\tag{5}
 + f_i \partial^\mu \left( - \frac{1}{2} \mathcal{D}_\mu^{ij} d_j\right) + 
\overline{\xi}\partial^\mu \left( \frac{i}{4}\gamma^j \psi_\mu d_j - \frac{1}{2} D_\mu \zeta \right)
\end{equation}
where $c_i$, $e_i$, $d_i$, $f_i$ are FD while $\zeta ,\xi$ are Majorana BE spinors.

The effective action $S_{c1} + S_{gf} + S_{FP}$ possesses the global Becci-Rouet-Stora-Tyutin (BRST) invariance [3, 4, 5]
\begin{equation}\tag{6a}
\delta b_N^i = \left(-\mathcal{D}_\mu^{ij} c_j + \frac{1}{2} \epsilon^{ipj} b_{\mu p} d_j + \frac{i}{4}\overline{\psi}_\mu \gamma_i\zeta\right)\eta
\end{equation}
\begin{equation}\tag{6b}
\hspace{-4.8cm}\delta w_\mu^i = -\frac{1}{2} \mathcal{D}_\mu^{ij} d_j \eta
\end{equation}
\begin{equation}\tag{6c}
\hspace{-2cm}\delta \psi_\mu = \left(\frac{i}{4}\gamma_j \psi_\mu d^j - \frac{1}{2}   D_\mu \zeta\right)\eta
\end{equation}
\begin{equation}\tag{7a}
\hspace{-4.7cm}\delta e^i = - \eta N_b^i
\end{equation}
\begin{equation}\tag{7b}
\hspace{-4.7cm}\delta f^i = - \eta N_w^i
\end{equation}
\begin{equation}\tag{7c}
\hspace{-4.7cm}\delta \overline{\xi} = - \eta \overline{N}_\psi
\end{equation}
\begin{equation}\tag{8a}
\hspace{-2cm}\delta c_i = \frac{1}{2}\epsilon_{ijk} c^jd^k \eta -  \frac{i}{16} \overline{\zeta}  \gamma_i \zeta\eta
\end{equation}
\begin{equation}\tag{8b}
\hspace{-4cm}\delta d_i = \frac{1}{4}\epsilon_{ijk} d^jd^k \eta 
\end{equation}
\begin{equation}\tag{8c}
\hspace{-1cm}\delta \zeta =  \frac{i}{4}\gamma_i d^i\zeta \eta\qquad \left(\delta\overline{\zeta} =   \frac{-i}{4} \;\overline{\zeta} \gamma_i d^i \eta\right)
\end{equation}
where $\eta$ is a FD constant scalar.  If the variation of a field $\Phi$ is of the form $\delta\eta$ where $\delta$ is a right variation, then $\delta^2 = 0$.  For example, we have
\begin{align}\tag{9}
\delta^2c_i &= \delta \left[ \frac{1}{2} \epsilon_{ijk} c^jd^k - \frac{i}{16}\overline{\zeta}\gamma_i \zeta \right]\\
&= \frac{1}{2} \epsilon_{ijk}c^j \left(\frac{1}{4} \epsilon^{k\ell m} d_\ell d_m\right) - \frac{1}{2} \epsilon_{ijk} \Big( \frac{1}{2} \epsilon^{j\ell m} c_\ell d_m\nonumber \\
&\hspace{2cm}- \frac{i}{16} \overline{\zeta} \gamma^j\zeta \Big) d^k - \frac{i}{16}\Bigg[ \overline{\zeta} \gamma_i \left( \frac{i}{4} \gamma^jd_j\right)\zeta\nonumber \\
&\hspace{3cm}+ \left( -\frac{i}{4} \overline{\zeta} \gamma_j d^j\right) \gamma_i \zeta\Bigg]\nonumber \\
& = 0.\nonumber
\end{align}
Consequently, the BRST transformation is idempotent.

We will now consider radiative corrections in this model.  Of particular interest if the phase associated with the one loop effective action associated with the field $w_\mu^i$ having a background contribution $\Omega_\mu^i$.  (We are replacing $w_\mu^i$ by $\Omega_\mu^i + \sigma_\mu^i$ where $\sigma_\mu^i$ is a quantum fluctuation.)  The gauge invariances of eq. (2) in the background field $\Omega_\mu^i$ can be preserved if the gauge fixing of eq. (3) gets replaced by [6, 7]
\begin{equation}\tag{10a-c}
\mathcal{D}_\mu^{ij}(\Omega)b_j^\mu = \mathcal{D}_\mu^{ij}(\Omega)\sigma_j^\mu
= D_\mu(\Omega) \psi^\mu = 0
\end{equation}
where
\begin{equation}\tag{11a}
\mathcal{D}_\mu^{ij} (\Omega) \equiv \partial_\mu \eta^{ij} - \epsilon^{ipj} \Omega_{\mu p}
\end{equation}
\begin{equation}\tag{11b}
D_\mu (\Omega) \equiv \partial_\mu + \frac{i}{2} \gamma^j \Omega_{\mu j}.
\end{equation}
The ghost action of eq. (5) now is replaced by
\begin{align}\tag{12}
\mathcal{L}_{FP} &= e_i \mathcal{D}^{\mu ij}(\Omega) \left[ - \mathcal{D}_{\mu jk} (\Omega + \sigma) c^k + \frac{1}{2} \epsilon_{jpk} b_\mu^p + \frac{1}{4} \overline{\psi}_\mu\gamma_j \zeta \right]\\
& + f_i \mathcal{D}^{\mu ij}(\Omega) \left[ - \frac{1}{2}\mathcal{D}_{\mu jk} (\Omega + \sigma) d^k \right] + \overline{\xi}D^\mu (\Omega ) \left[ \frac{i}{4} \gamma^j \psi_\mu d_j -  \frac{1}{2}  D_\mu (\Omega + \sigma) \zeta \right]\nonumber
\end{align}
as the gauge transformation in which the background field $\Omega_{\mu i}$ is unaltered is broken.

The terms in $\mathcal{L}_{c1} + \mathcal{L}_{gf}$ in the gauge in which $\alpha_b = \alpha_w = \alpha_\psi = 0$ that are bilinear in the fields $\Phi^T = \left( b_\mu^i, w_\mu^i, \psi_\mu, N_b^i, N_w^i, N_\psi\right)^T$ and 
 $\overline{\Phi} = \left( b_\mu^i, w_\mu^i, \overline{\psi}_\mu, N_b^i, N_w^i, \overline{N}_\psi\right)$ are of the form
\begin{equation}\tag{13}
\overline{\Phi} H_\kappa \Phi
\end{equation}
where
\begin{equation}\tag{14}
H = \left(
\begin{array}{cccccc}
0 & \mathcal{D}_\lambda^{ij}\epsilon^{\mu\lambda\nu} & 0 & \mathcal{D}_\mu^{ij} & 0 & 0 \\
\mathcal{D}_\lambda^{ij}\epsilon^{\mu\lambda\nu} & 0 & 0 & 0 & \mathcal{D}_\mu^{ij} & 0\\
0 & 0 & D_\lambda\epsilon^{\mu\lambda\nu} & 0 & 0 & D_\mu\\
-\mathcal{D}_\nu^{ij} & 0 & 0 & 0 & 0 & 0\\
0 & -\mathcal{D}_\nu^{ij} & 0 & 0 & 0 & 0 \\
0 & 0 & -D_\nu & 0 & 0 & 0
\end{array}
\right)
\end{equation}
where the derivatives appearing in eq. (14) are those of eq. (11).

One-loop effects in this model are given by $\det H$, but we must deal with $\det^{1/2}H^2$.  However, if $H$ has negative eigenvalues, one must separately compute the phase $\Theta$ of $\det H$.  This is determined by [8-11]
\begin{equation}\tag{15}
\Theta = - \frac{\pi}{2}\lim\limits_{s\rightarrow 0} \eta(s)
\end{equation}
where $\eta(s) = \eta_{\lambda=1}(s)$ with
\begin{equation}\tag{16}
\frac{d\eta_\lambda(s)}{d\lambda}= - \frac{s}{\Gamma\left(\frac{s+1}{2}\right)} 
\int_0^\infty dt \,t^{\frac{s-1}{2}} str \left[ \frac{dH_\lambda}{d\lambda} e^{-H_\lambda^2t} \right].
\end{equation}
In eq. (16), $H_\lambda$ is obtained from $H$ by rescaling each of the external fields by a factor of $\lambda$ [16].  It follows  from eq. (14) that 
\begin{equation}
\hspace{-4cm}H = \left(
\begin{array}{ccc}
-\eta_{\mu\nu}\mathcal{D}^{2ij} -\epsilon^{ijp}R_{\mu\nu p} & 0 & 0 \\
0 & -\eta_{\mu\nu}\mathcal{D}^{2ij}-\epsilon^{ijp}R_{\mu\nu p} & 0 \\
0 & 0 &  -\eta_{\mu\nu}D^2- \frac{i}{2}\gamma \cdot R_{\mu\nu } \\
0 & -\frac{1}{2}\epsilon_{\lambda\sigma\nu}\epsilon^{ijp}R^{\lambda\sigma}_p & 0  \\
-\frac{1}{2}\epsilon_{\lambda\sigma\nu}\epsilon^{ijp}R^{\lambda\sigma}_p &0 & 0  \\
0 & 0 & -\frac{1}{2}\epsilon_{\lambda\sigma\nu}\left(\frac{i}{2} \gamma \cdot R^{\lambda\sigma}\right)
\end{array}\right.\nonumber
\end{equation}
\begin{equation}\tag{17}
\hspace{4.5cm}\left. \begin{array}{ccc}
0 & \frac{1}{2}\epsilon_{\mu\lambda\sigma}\epsilon^{ijp}R^{\lambda\sigma}_p & 0\\
\frac{1}{2}\epsilon_{\mu\lambda\sigma}\epsilon^{ijp}R^{\lambda\sigma}_p & 0 &  0\\
0 & 0 & \frac{1}{2}\epsilon_{\mu\lambda\sigma}\left( \frac{i}{2} \gamma \cdot R^{\lambda\sigma}\right)\\
 -\mathcal{D}^{2ij}& 0 & 0\\
0 & -\mathcal{D}^{2ij} & 0\\
0 & 0 & -D^2
\end{array}\right)
\end{equation}
Since to compute $\Theta$ in eq. (15) we only need $\eta(s)$ with $s = 0$, it follows that in eq. (15) the only contribution that is needed in the integrand in the integral over $t$ is the piece that results in a pole at $s = 0$.  We now follow the approach to compution $\Theta$ used in ref. [11]. If $e^{-H^2_{\lambda}t}$ is expanded in powers of $t$, then this entails using terms that behave as $t^0$ or $t^1$.  Since $H_{\lambda}^2$ is of the form $(p + V)^2 + \phi (p \equiv -i\partial)$, it follows [8] that the term linear in $t$ is the Seeley-Gilkey coefficient $a_1 = -\phi$ and so what is needed in $str\left(\frac{dH_\lambda}{d\lambda} e^{-H_\lambda^2t}\right)$ is the contribution that is linear in $t$, which is
\begin{equation}
str
\begin{pmatrix}
0 & -\epsilon^{ipj}\epsilon^{\mu\lambda\nu}\Omega_{\lambda p} & 0 & -\epsilon^{ipj}\Omega_{\mu p} & 0 & 0\\
-\epsilon^{ipj}\epsilon^{\mu\lambda\nu}\Omega_{\lambda p} & 0 & 0 & 0 &
-\epsilon^{ipj}\Omega_{\mu p} & 0\\
0 & 0 & \frac{i}{2}\gamma \cdot \Omega_\lambda\epsilon^{\mu\lambda\nu} & 0 & 0 & 
\frac{i}{2}\gamma \cdot  \Omega_\mu\\
\epsilon^{ipj}\Omega_{\nu p} & 0 & 0 & 0 & 0 & 0 \\
0 & \epsilon^{ipj}\Omega_{\nu p} & 0 & 0 & 0 & 0  \\
0 & 0 & -\frac{i}{2}\gamma \cdot \Omega_\nu & 0 & 0 & 0
\end{pmatrix}\nonumber
\end{equation}
\begin{equation}
\times
\begin{pmatrix}
-\lambda \epsilon^{ijp}R_{\mu\nu p} & 0 &  0 & 0 & \frac{1}{2}\lambda \epsilon_{\mu\lambda\sigma}\epsilon^{ijp}R^{\lambda\sigma}_p & 0\\
0 & -\lambda \epsilon^{ijp}R_{\mu\nu p} & 0 &\frac{1}{2}\lambda \epsilon_{\mu\lambda\sigma}\epsilon^{ijp}R^{\lambda\sigma}_p & 0 & 0 \\
0 & 0 & -\frac{i\lambda}{2}\gamma \cdot R_{\mu\nu} & 0 & 0 & \frac{\lambda}{2} \epsilon_{\lambda\sigma\nu} \left(\frac{i}{2}\gamma\cdot R^{\lambda\sigma}\right)\\
0 & -\frac{1}{2}\lambda \epsilon_{\lambda\sigma\nu}\epsilon^{ijp}R^{\lambda\sigma}_p & 0 & 0 & 0 & 0\\
-\frac{1}{2}\lambda \epsilon_{\lambda\sigma\nu}\epsilon^{ijp}R^{\lambda\sigma}_p & 0 & 0 & 0 & 0 & 0\\
0 & 0 & -\frac{\lambda}{2} \epsilon_{\lambda\sigma\nu} \left(\frac{i}{2}\gamma\cdot R^{\lambda\sigma}\right) & 0 & 0 & 0
\end{pmatrix}\nonumber
\end{equation}
\begin{align}\tag{18}
&= - tr  \Bigg[\left( \frac{i}{2} \gamma \cdot \Omega_\lambda \epsilon^{\mu\lambda\nu}\right) \left(- \frac{i}{2 }\lambda\gamma \cdot R_{\mu\nu}\right) + \left( \frac{i}{2}\gamma \cdot \Omega_{\nu}\right)
\left( -\frac{\lambda}{2} \epsilon_{\lambda\sigma\nu}\right)\left( \frac{i}{2} \gamma \cdot R^{\lambda\sigma}\right) \nonumber \\
&\hspace{3cm}+ \left( -\frac{i}{2} \gamma \cdot \Omega_\nu\right) \left( \frac{\lambda}{2}\epsilon_{\lambda\sigma\nu}\right)\left(\frac{i}{2} \gamma \cdot R^{\lambda\sigma}\right) \Bigg]\nonumber \\
& = 0. \nonumber
\end{align}
This shows that the phase of the one-loop effective action vanishes non-trivially.  This is unlike what happens when a gauge field couples to spinor in $2 + 1$ dimensions; in this case the phase of the one-loop effective action contributes a Chern-Simons term which means that the gauge field develops a topological mass and that the gauge coupling is quantized [11].  Other radiative effects in the model of eq. (1) can be computed using operator regularization [11, 12]. This technique is especially useful as when using it, the initial action is unaltered, thereby leaving the symmetries of eqs. (2, 6, 7, 8) intact\vspace{.6cm}.

\noindent
{\Large\bf{Acknowledgements}}\\
 F.T. Brandt and T.N. Sherry had many helpful suggestions. R. Macleod made had a useful comment.

\section*{Appendix I - Notation}

We use the flat space metric
\begin{equation}\tag{A.1}
\eta^{ij} = \mathrm{diag} (+, -, -)
\end{equation}
with the anti-symmetric tensor
\begin{equation}\tag{A.2}
\epsilon^{012} = +1 = \epsilon_{012}.
\end{equation}
The Dirac matrices are
\begin{equation}\tag{A.3}
\gamma^0 = \sigma_2, \quad \gamma^i = i\sigma_3, \quad \gamma^2 = i\sigma_1
\end{equation}
so that
\begin{equation}\tag{A.4}
\gamma^i \gamma^j = \eta^{ij} + i\epsilon^{ijk}\gamma_k.
\end{equation}
The Majorana condition is $\psi = \psi_C$ where
\begin{equation}\tag{A.5}
\psi_C = C \overline{\psi}^T
\end{equation}
with
\begin{equation}\tag{A.6}
\overline{\psi} = \psi^\dagger \gamma^0
\end{equation}
and $C = - \gamma^0 = C^{-1}$ so that
\begin{equation}\tag{A.7}
C \gamma^i C^{-1} = - \gamma^{iT} = \gamma^{i\dagger}.
\end{equation}
With our conventions, if $\psi = \psi_C$ then $\psi = \psi^*$.  If $\phi$ and $\chi$ are Grassmann spinors then
\begin{equation}\tag{A.8a-c}
\overline{\chi} \phi = \overline{\phi}\chi , \quad
\overline{\chi} \gamma^i \phi = - \overline{\phi}\gamma^i \chi , \quad
\overline{\chi} \gamma^i \gamma^j \phi =  \overline{\phi}\gamma^j\gamma^i \chi.
\end{equation}
The signs in eq. (A.8) are reversed if either or both $\chi$ and $\phi$ are not Grassmann.

We also have defined
\begin{equation}\tag{A.9a}
R_{\mu\nu}^i = \partial_\mu w_\nu^i - \partial_\nu w_\mu^i - \epsilon^{ijk}w_{\mu j} w_{\nu k}
\end{equation}
\begin{equation}\tag{A.9b}
D_\mu = \partial_\mu + \frac{i}{2} \gamma^i w_{\mu i}  \quad \left( \left[ D_\mu, D_\nu \right] = \frac{i}{2}\gamma^i R_{\mu \nu i} \right)
\end{equation}
\begin{equation}\tag{A.9c}
\mathcal{D}_\mu^{ij} = \partial_\mu -\epsilon^{ipj} w_{\mu p}  \quad \left( \left[ \mathcal{D}_\mu, \mathcal{D}_\nu \right]^{ij} = \epsilon^{ijk} R_{\mu\nu k}\right)
\end{equation}

\section*{Appendix II - The First Order Einstein-Hilbert Action in $1 + 1$ Dimensions}

In this paper we have demonstrated how a BRST effective action can be derived for a theory with an unusual gauge invariance by considering supergravity in $2 + 1$ dimensions.  In this appendix we further illustrate this by deriving the effective action for the first order Einstein-Hilbert (Palatini) action in $1 + 1$ dimensions.  This theory is manifestly invariant under a diffeormorphism transformation, but it is not this gauge transformation that follows from the first class constraints in the theory [13].

If in the Einstein-Hilbert action
\begin{equation}\tag{B.1}
S_{c1} = \int d^d x \sqrt{-g}\; g^{\mu\nu} R_{\mu\nu} (\Gamma)
\end{equation}
we set
\begin{equation}\tag{B.2a}
h^{\mu\nu} = \sqrt{-g}\; g^{\mu\nu}
\end{equation}
\begin{equation}\tag{B.2b}
G^\lambda_{\mu\nu} = \Gamma_{\mu\nu}^\lambda - \frac{1}{2} \left( \delta_\mu^\lambda  \Gamma_{\nu\sigma}^\sigma + \delta_\nu^\lambda \Gamma_{\mu\sigma}^\sigma\right)
\end{equation}
then
\begin{equation}\tag{B.3}
S_{c1} = \int d^d x h^{\mu\nu} \left( G_{\mu\nu , \lambda}^\lambda + \frac{1}{d - 1} G_{\lambda\mu}^\lambda G_{\sigma\nu}^\sigma - G_{\sigma\mu}^\lambda G_{\lambda\nu}^\sigma \right).
\end{equation}
When $d = 2$, the first class constraints show that in addition to diffeomorphism invariance, the action of eq. (B.3) has the local gauge invariance [13]
\begin{equation}\tag{B.4a}
\delta h^{\mu\nu} = \left( \epsilon^{\mu\lambda} h^{\sigma\nu} + \epsilon^{\nu\lambda} h^{\sigma\mu} \right) \theta_{\lambda\sigma}
\end{equation}
\begin{equation}\tag{B.4b}
\delta G_{\mu\nu}^\lambda = - \epsilon^{\lambda\rho} \theta_{\mu\nu , \rho} - \epsilon^{\rho\sigma} 
\left( G_{\mu\rho}^\lambda \theta_{\sigma\nu} +  G_{\nu\rho}^\lambda \theta_{\sigma\mu} \right)
\end{equation}
where $\epsilon^{01} = 1 = -\epsilon_{01} = -\epsilon^{10}$ and $\theta^{\mu\nu} = \theta^{\nu \mu}$.

It is not clear how to extend the gauge invariance of eq. (B.4) to $d > 2$ or if it is possible to couple matter fields to $h^{\mu\nu}$, $G_{\mu\nu}^\lambda$ so that these symmetries are maintained.

The gauge fixing condition [14]
\begin{equation}\tag{B.5}
\epsilon_{\lambda\sigma} G_{\mu\nu}^{\lambda ,\sigma} = 0
\end{equation}
results in the gauge fixing Lagrangian
\begin{equation}\tag{B.6}
\mathcal{L}_{gf} = \frac{\alpha}{2} N_{\mu\nu} N^{\mu\nu} - N^{\mu\nu}\epsilon_{\lambda\sigma} G_{\mu\nu}^{\lambda ,\sigma}
\end{equation}
as well the Faddeev-Popov ghost action
\begin{equation}\tag{B.7}
\mathcal{L}_{gh} = -\overline{\zeta}^{\alpha\beta} \epsilon_{\mu\nu} 
\left[  -\epsilon^{\mu\rho} \zeta_{\alpha\beta ,\rho} - \
\epsilon^{\rho\sigma}\left(  G_{\alpha \rho}^\mu \zeta_{\sigma\beta} + 
G_{\beta\rho}^\mu \zeta_{\sigma\alpha}\right)\right]^{, \nu}
\end{equation}
where $\zeta_{\mu\nu},\overline{\zeta}^{\mu\nu}$ are a pair of symmetric tensor Grassmann fields.  The effective action $S_{c1} + S_{gf} + S_{gh}$ has the unusual global BRST gauge invariance
\begin{equation}\tag{B.8a}
\hspace{-2.7cm}\delta h^{\mu\nu} = (\epsilon^{\mu\lambda} h^{\sigma\nu} + \epsilon^{\nu\lambda} h^{\sigma\mu}) \zeta_{\lambda\sigma}\eta
\end{equation}
\begin{equation}\tag{B.8b}
\delta G_{\mu\nu}^\lambda = -\epsilon^{\lambda\rho} \zeta_{\mu\nu , \rho} \eta -\epsilon^{\rho\sigma} \left( G_{\mu\rho}^\lambda \zeta_{\sigma\nu} + G_{\nu\rho}^\lambda \zeta_{\sigma\mu}\right)\eta
\end{equation}
\begin{equation}
\hspace{-5.3cm}\delta \overline{\zeta}^{\alpha\beta} = N^{\alpha\beta} \eta\nonumber
\end{equation}
\begin{equation}\tag{B.8c}
\hspace{-2cm}\delta \zeta_{\alpha\beta} = \frac{1}{2} \epsilon^{\mu\nu} \left( \zeta_{\mu\alpha} \zeta_{\nu\beta} + \zeta_{\nu\alpha} \zeta_{\mu\beta} \right) \eta
\end{equation}
where $\eta$ is a constant Grassmann scalar.

Radiative corrections arising in this model have been discussed in ref. [14].

We note that if the Hamiltonian approach the finding a BRST invariant action, the resulting effective action itself possesses a gauge invariance [15].

\end{document}